\documentclass[12pt]{article}

\usepackage{amsmath,amssymb,amscd}

\usepackage{graphicx}% Include figure files

\newcommand{\vol}[1]{\textbf{#1},   }
\newcommand{\eprint}[1]{\texttt{#1}}

\newcommand{\tref}[1]{(\ref{#1})}
\newcommand{\href}[2]{#2}

\newcommand{\bea}{\begin{eqnarray}}
\newcommand{\eea}{\end{eqnarray}}
\newcommand{\beq}{\begin{equation}}
\newcommand{\eeq}{\end{equation}}
\newcommand{\nnel}{\nonumber \\ {}}

\newcommand{\kav}{\langle k \rangle}

\begin{document}
\renewcommand{\thefootnote}{\fnsymbol{footnote}}

\begin{flushright}
 \texttt{Imperial/TP/06/TSE/1} \\
 \eprint{cond-mat/0607196} \\
 Final version 21st July 2006\\
 Eur.\ Phys. J. B \vol{56} (2007) 65-69\\
 DOI: \texttt{10.1140/epjb/e2007-00084-8}\\
 Original publication available at\\
 \href{http://www.springerlink.com/content/l3j16569hu1g1813}{\texttt{www.europhysj.org}}
 \end{flushright}

\begin{center}
{\Large\textbf{Exact Solutions for Network Rewiring Models}} \\[0.5cm]
 {\large T.S.\ Evans\footnote{\href{http://www.imperial.ac.uk/people/t.evans}{\texttt{http://www.imperial.ac.uk/people/t.evans}} }},
 \\[0.5cm]
 \href{http://www.imperial.ac.uk/research/theory}{Theoretical Physics},
 Blackett Laboratory, Imperial College London,\\
 South Kensington campus, London, SW7 2AZ,  U.K.
\end{center}

%\date{23rd April 2006. Revised version 7th July. Minor corrections 21st July}

\abstract{Evolving networks with a constant number of edges may be
modelled using a rewiring process. These models are used to describe
many real-world processes including the evolution of cultural
artifacts such as family names, the evolution of gene variations,
and the popularity of strategies in simple econophysics models such
as the minority game. The model is closely related to Urn models
used for glasses, quantum gravity and wealth distributions. The full
mean field equation for the degree distribution is found and its
exact solution and generating solution are given.}

% PACS, the Physics and Astronomy Classification Scheme.
%\pacs{89.75.Da, 89.65.Ef, 89.65.Gh, 89.75.Kd}
%\PACS{
% {89.75.Hc}{Networks and genealogical trees} \and
% {89.75.Da}{Systems obeying scaling laws} \and
% {89.65.Ef}{Social organizations; anthropology} \and
% {89.65.Gh}{Economics; econophysics, financial markets, business and
% management}
% }

 % 89.75.Hc    Networks and genealogical trees

%\maketitle

\renewcommand{\thefootnote}{\arabic{footnote}}
\setcounter{footnote}{0}

% *******************************************************

\vspace*{2cm}

Networks with a constant number of edges that evolve only through
a rewiring of those edges are of great importance, as exemplified
by Watts and Strogatz \cite{WS98}. Many different applications may
be modelled as a network rewiring: the transmission of cultural
artifacts such as pottery designs, dog breed and baby name
popularity \cite{Neiman95,BSHB03,HBH04,BHS04}, the distribution of
family names in constant populations \cite{ZM01}, the diversity of
genes \cite{KC64,CK70} and the popularity of minority game
strategies \cite{ATBK04}.  There is a close link to some models of
the zero range process \cite{EH05} and the closely related Urn
type-models used for glasses \cite{Ritort95,BBJ97}, simplicial
gravity \cite{BBJ99} and wealth distributions \cite{BDJKNPZ02}.
The rewiring of networks is also studied in its own right
\cite{WS98,PLYXZW,OHTY}.

However previous analytic results for network rewiring models are
based on incomplete mean field equations and their approximate
solutions. In this letter I give the full equations for linear
removal and attachment probabilities with their exact solution. This
means the analytic results for rewiring models can match the status
of those for random graph and growing network models (e.g.\ see
\cite{DM01KR01}).

\begin{figure}
{\centerline{\includegraphics[width=6cm]{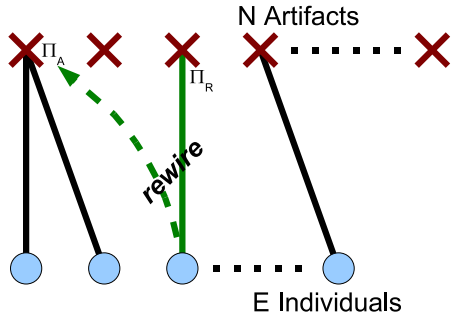}}}
\caption{In the abstract form of the model, there are $E$ vertices
of one type---`individual' vertices. Each has one edge which runs to
one of $N$ vertices of a second type---the `artifacts'. The degree
of the artifact vertices is $k$ indicating that one artifact has
been chosen by $k$ distinct individuals. The rewiring will be of the
artifact ends of the edges, so each individual always has one edge.}
\label{fig:CopyModel}
\end{figure}
Consider the degree distribution of the artifact
vertices\footnote{The artifacts may be a dog breed, baby name or
pottery style with each individual choosing one type of artifact as
indicated by its edge \cite{Neiman95,BSHB03,HBH04,BHS04}.  For
family names the individuals are those who inherit the name from
their partner, the edges are the partners who retain their family
name while the artifacts represent different family names. For a
model of gene distributions \cite{KC64,CK70} in a haploid
population, the artifacts are the alleles while the individuals are
the organisms. A diploid population may be modelled in a similar
manner. In Urn/Backgammon/Balls-in-Boxes models
\cite{Ritort95,BBJ97,BBJ99,BDJKNPZ02} the individuals represent the
balls while the artifacts are the boxes. The rewiring of an
undirected network is described by the same equations provided $E$
is replaced by $2E$.}, $n(k)$, in the bipartite graph of Fig.\
\ref{fig:CopyModel}. At each time step I make two choices then alter
the network.

First I choose one of the $E$ individuals at random\footnote{In this
letter `random' without further qualification indicates that a
uniform distribution is used to draw from the set under
discussion.}, aiming to rewire the artifact end of the chosen
individual's one edge. Thus the edge to be rewired is from an
artifact chosen by `preferential removal'.

Second, this edge will be reattached to one of the $N$ artifact
vertices chosen with probability $\Pi_A$. With probability $p_p$
preferential attachment is used to choose the artifact. This is
equivalent to choosing an individual at random and \emph{copying}
the current artifact choice of that individual. Alternatively with
probability $p_r$ an artifact is chosen at random to receive the
rewired edge. This corresponds to \emph{innovation} in the context
of cultural transmission \cite{Neiman95,BHS04}, in gene evolution it
is \emph{mutation} \cite{KC64,CK70}.

With only these types of event, $p_p+p_r=1$, the number of artifacts
$N$ is constant, and
\beq
 \Pi_R = \frac{k}{E}, \qquad
 \Pi_A = p_r\frac{1}{N} + p_p\frac{k}{E},
   \qquad (0 \leq k \leq E) .
 \label{PiRPiAsimple}
\eeq
After these choices have been made the rewiring takes place.  The
mean field equation for the degree distribution for $(0 \leq k \leq
E)$ is therefore
\begin{eqnarray}
\lefteqn{n(k,t+1) - n(k,t) =}
\nnel
 &&  n(k+1,t) \Pi_R(k+1) \left( 1- \Pi_A(k+1) \right)
 \nnel
& -& n(k,t)   \Pi_R(k)   \left( 1- \Pi_A(k)   \right)
     - n(k,t)   \Pi_A(k)   \left( 1- \Pi_R(k)   \right)
   \nnel
 &+& n(k-1,t) \Pi_A(k-1) \left( 1- \Pi_R(k-1) \right)
   .
   \label{neqngen}
\end{eqnarray}
This equation holds at the boundaries $k=0$ and $k=E$ provided
$n(k)=\Pi_R(k)=\Pi_A(k)=0$ for $k=-1$ and $k=(E+1)$ are chosen. Note
that by including the factors of $(1-\Pi_A)$ and $(1-\Pi_R)$ on the
right hand side I am explicitly excluding events where the same
vertex is chosen for removal and attachment in any one rewiring
event since such events do not change the distribution.  More
importantly these terms ensure a rigid upper boundary $n(k>E,t)=0$.
Contrast this with the equations for growing networks (for instance
in \cite{DM01KR01}) where there is no rigid upper boundary in the
long time limit and such terms are absent . These additional factors
are only significant for $k \sim E$ but they are missing in other
discussions of rewiring models. The condition $E \gg k$ is usually
sufficient for these factors to be negligible and results match the
literature in this regime.

These equations \tref{neqngen} have a stationary solution if
\bea
 \lefteqn{n(k+1) \Pi_R(k+1) \left( 1- \Pi_A(k+1) \right)}
 \nnel
 &=&
 n(k)   \Pi_A(k)   \left( 1- \Pi_R(k)   \right)
 \eea
 which gives the static solution
$n(k)$ for $(E \geq k \geq 0)$ as
\bea
n(k) &=& A
 \frac{\Gamma( k + \widetilde{K} ) }
      {\Gamma (k+1 ) }
 \frac{\Gamma( E+\widetilde{E} - \widetilde{K}  - k) }
      {\Gamma (E+1  -  k  ) } ,
 \label{nsimpsol}
  \\
 && \widetilde{K} := \frac{p_r}{p_p}\kav, \;
\widetilde{E} := \frac{p_r}{p_p}E \; .
\eea
The normalisation constant $A$ can be found from \tref{GenFuncSol}
below and $\kav = E/N$ is the average artifact degree.

For $k \gg 1,\widetilde{K}$, the first ratio of Gamma functions
gives
\bea
\frac{\Gamma( k + \widetilde{K} ) }
      {\Gamma (k+1 ) } &\propto& k^{-\gamma} \left( 1+ O(\frac{1}{k},\frac{\widetilde{K}}{k}) \right),
 \label{gammasimp}
\eea
where $\gamma = 1- \widetilde{K} \leq 1$. This is consistent with
previous results which are usually given in a small mutation, $p_r
\approx 0$, and/or low average degree $\kav \ll 1$ limit.

The novel aspects in the present formulation are the extra factors
of $(1-\Pi_A)$ and $(1-\Pi_R)$ in \tref{neqngen}. These lead
directly to the second ratio of Gamma functions in \tref{nsimpsol}
which for $p_rE \gtrsim 1$ decays exponentially:
\bea
\frac{\Gamma\left(E+ \widetilde{E} - \widetilde{K} - k\right) }
      {\Gamma (E+1  -  k  ) }
 \propto  \exp \{ -\zeta k \} \left( 1+ O(\frac{k}{E})\right) \; ,
 \eea
where $\zeta = -\ln \left( p_p \right) +O(E^{-1})$.

However, when $p_r E \lesssim 1$ the numerator grows with $k$. In
fact at a critical random attachment probability, $p_r^*$, the total
distribution stops decreasing at the upper boundary, so $n(E) =
n(E-1)$. This occurs at
\bea
 p_r^* &=& \frac{E-1}{E^2+E(1-\kav)-1-\kav}
\eea
Therefore when $p_r < p_r^* \sim 1/E$ the degree distribution will
increase near $k=E$.

Thus there are two types of behaviour. For large innovation or
mutation, for $ 1 > p_r \gtrsim E^{-1}$ the distribution is
approximately $n(k) \propto (k)^{-\gamma} \exp\{-\zeta k\}$, a gamma
distribution, with an exact binomial distribution at $p_r = 1$, the
random graph case of \cite{WS98}.  This gives a power law for small
degree, $k \lesssim \ln(p_p^{-1})$, with an exponential cutoff for
higher degrees. Such behaviour is noted in the literature under
various approximations \cite{HBH04,BHS04,KC64,CK70,PLYXZW} and those
results are consistent the exact solution \tref{nsimpsol}. However
since $1 \ll \zeta$ implies $(\gamma -1) \ll \kav$, if one had only
one data set of a typical size, any power law section of reasonable
length ($k\lesssim \zeta^{-1}$) will have a power $\gamma$
indistinguishable from the value one (c.f.\ growing networks where
$\gamma >2$).

The second regime is where $p_r E \lesssim 1$, i.e.\ there is
usually no mutation or innovation over a time period when most edges
have been rewired once.  Here the tail of the distribution rises and
one artifact will be linked to almost all of the individuals. It is
the \emph{condensation} of \cite{BBJ97,BDJKNPZ02} and
\emph{fixation} in \cite{CK70} but again those results were given
only for the equivalent of large $E$.  Similar behaviour has been
discussed for growing networks, for example in \cite{DM01KR01}, but
not as an explicit network rewiring problem.

\begin{figure}
\includegraphics[width=7.8cm]{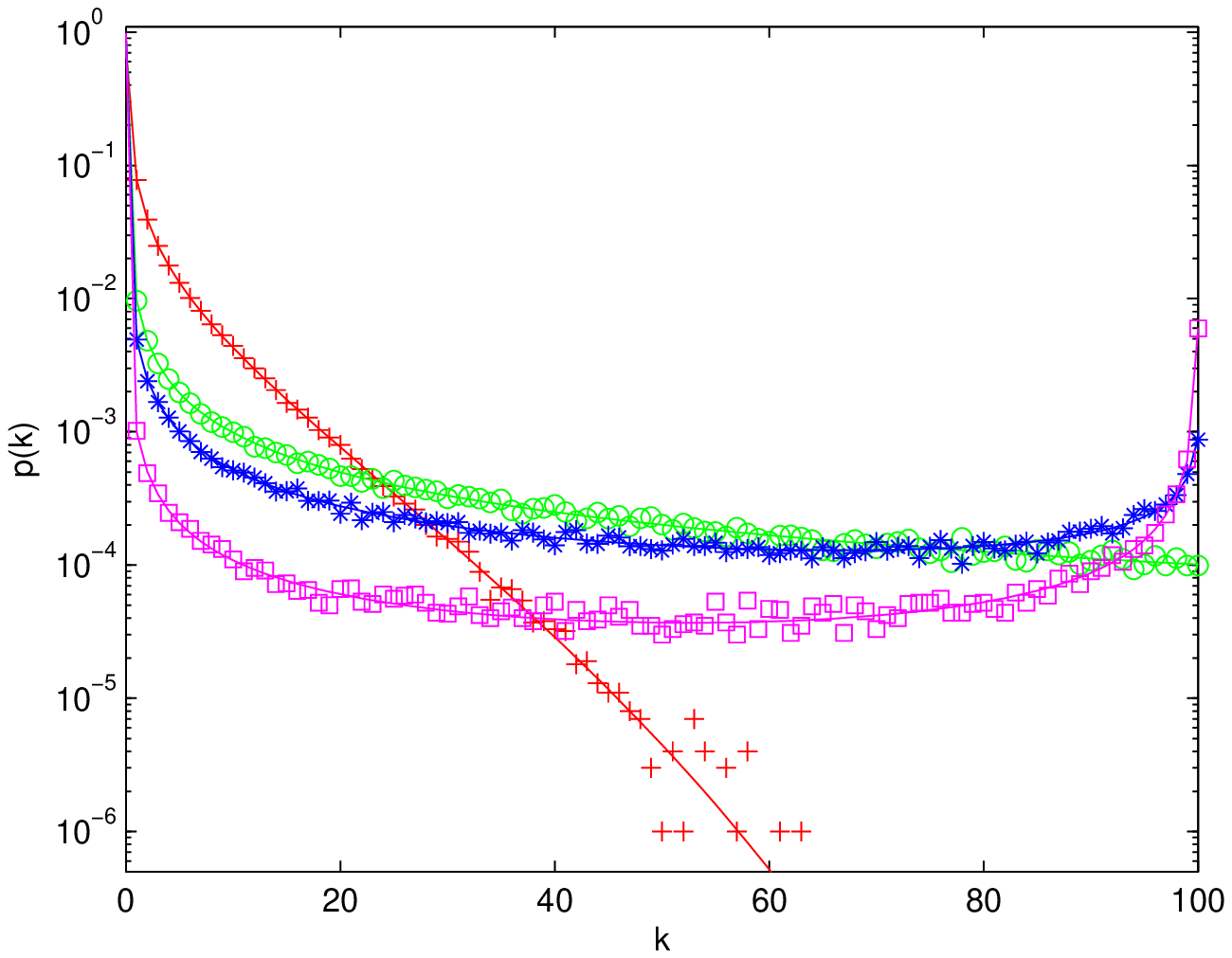}
\includegraphics[width=7.8cm]{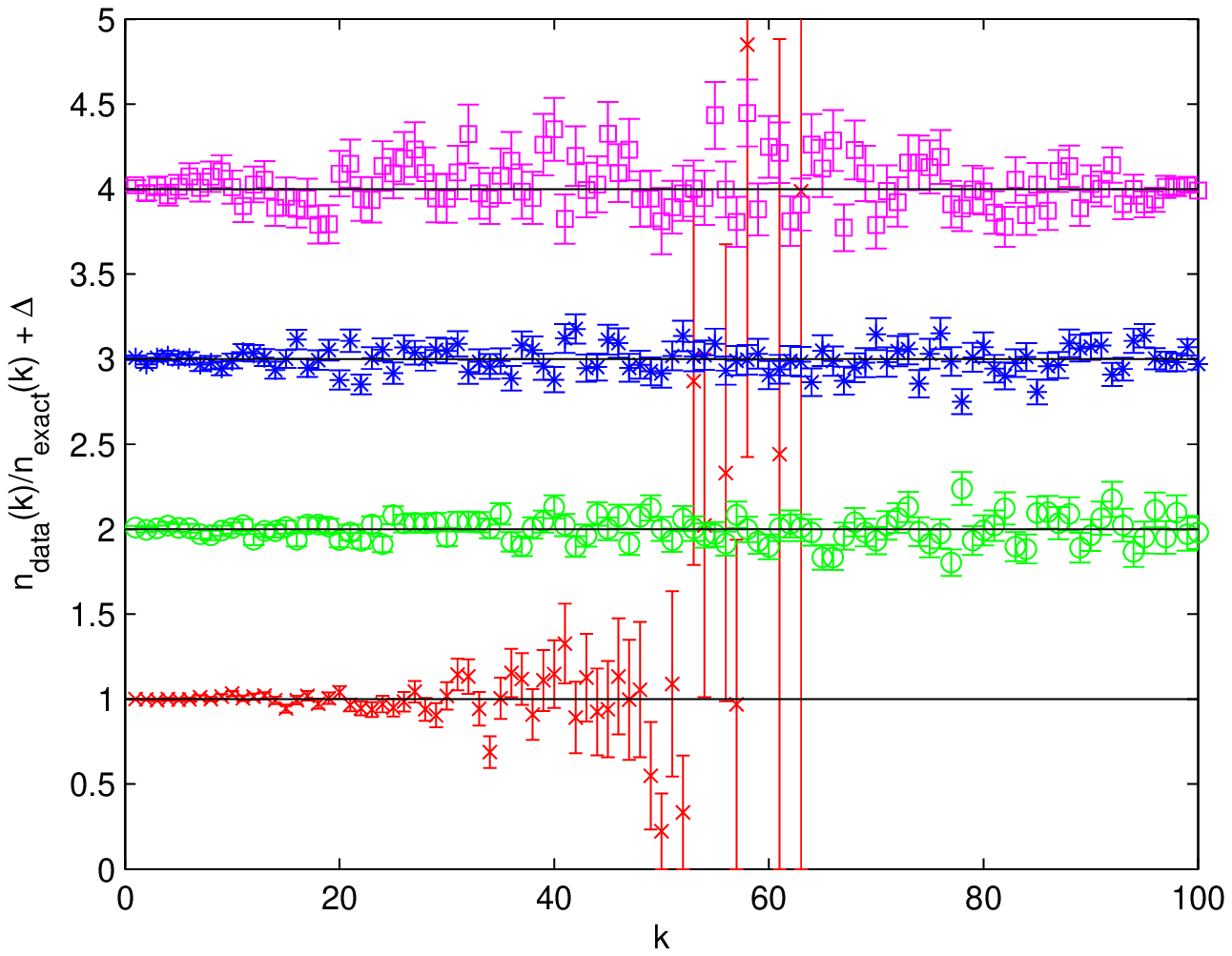}
\caption{Plots of the degree probability distribution function
$p(k)=n(k)/N$ and the fractional error shifted by $\Delta$ of the
data w.r.t.\ the exact solution.  For $N=E=100$ and various
$p_r=(1-p_p)=0.1$ (crosses, $\Delta=0$), $0.01$ (circles,
$\Delta=1$), $0.005$ (stars, $\Delta=2$) and $0.001$ (squares,
$\Delta=3$), while lines are the exact solutions. Measured after
$10^5$ rewiring events, averaged over $10^4$ runs. Started with
$n(k=1)=E$ but otherwise $n(k)=0$. The error bars are mostly smaller
or similar in size to the symbol in the first plot.}
\label{fig:PLYDD}
\end{figure}

\begin{figure}
\includegraphics[width=7.8cm]{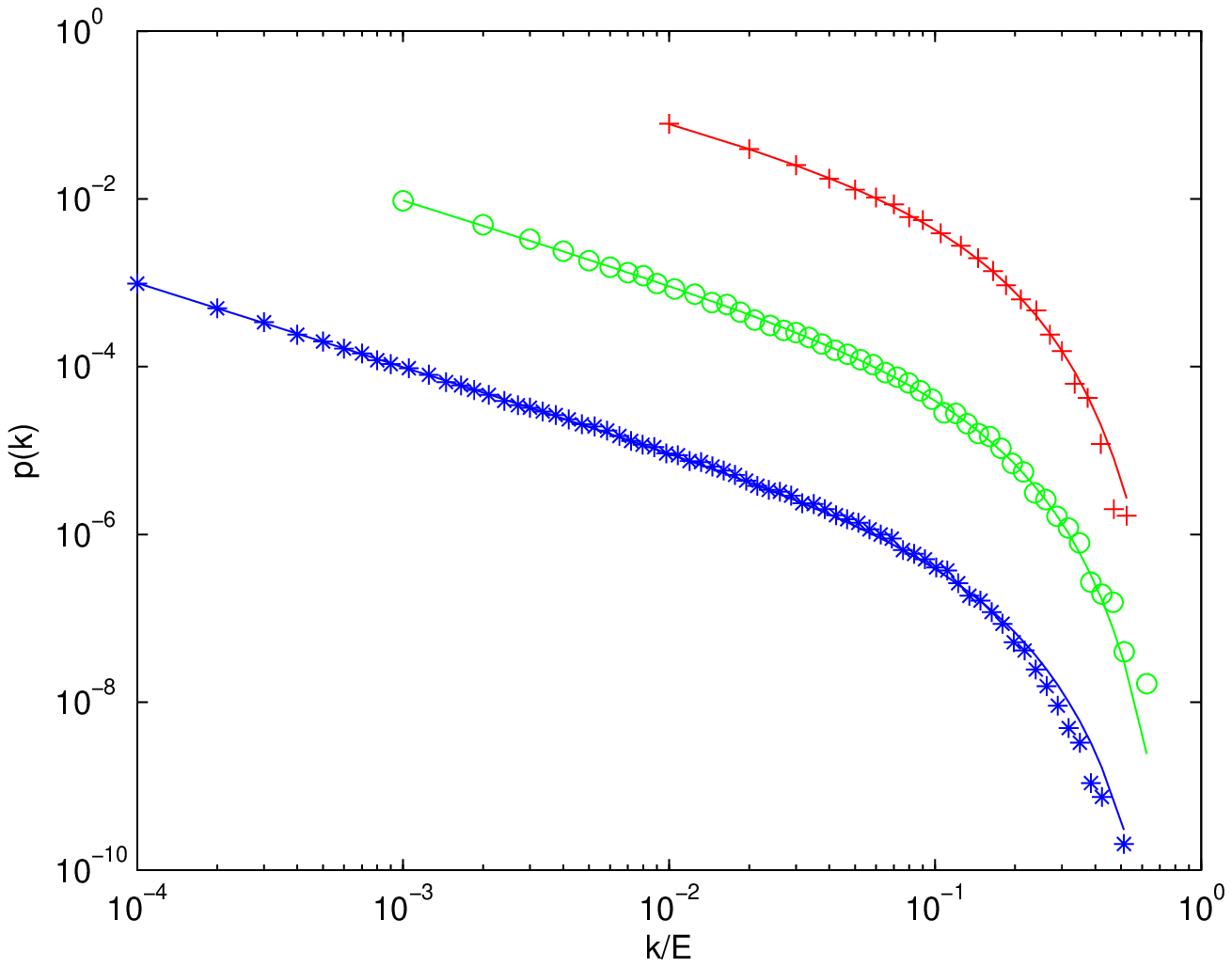}
\includegraphics[width=7.8cm]{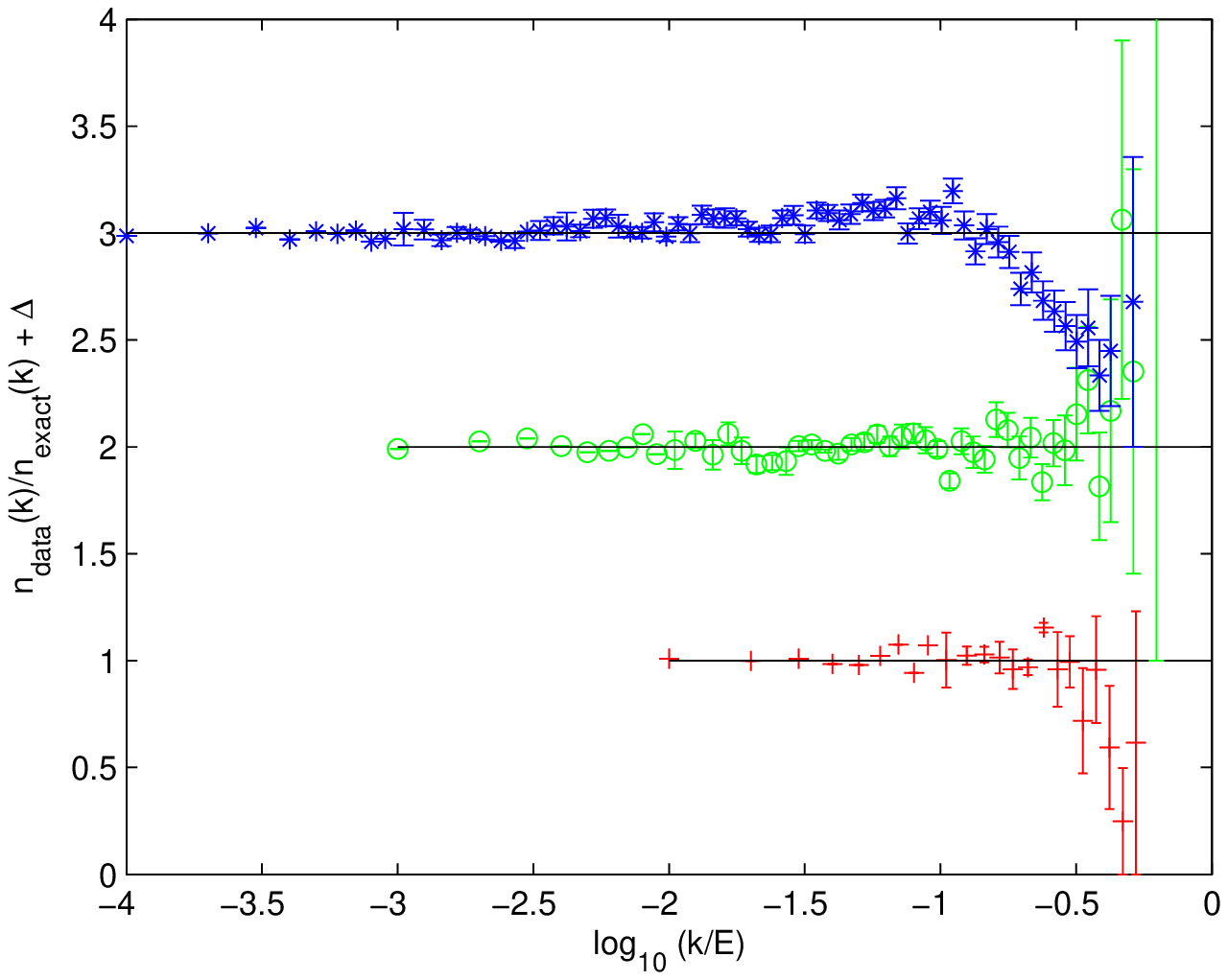}
\caption{The degree probability distribution function
 $p(k)=n(k)/N$ and the fractional error shifted by $\Delta$ w.r.t.\ the exact
solution for $N=E$, $Ep_r=10.0$ and $p_r=10^{-2}$ (crosses,
$\Delta=0$), $10^{-3}$ (circles, $\Delta=1$) and $10^{-4}$ (stars,
$\Delta=2$). Measured after $10^7$ rewiring events, averaged over
$10^3$ runs. Note that for $p_r=10^{-4}$ there are signs the model
may not have quite reached equilibrium. Started with $n(k=1)=E$ but
otherwise $n(k)=0$.} \label{fig:PLYDDEvary}
\end{figure}

In this simple model, there are no correlations between the degree
of vertices.  Indeed one need not impose a network structure as in
\cite{KC64,CK70,BBJ97,BBJ99,BDJKNPZ02}. Thus the mean-field
equations should be an excellent approximation to the actual
results. Numerical simulations confirm this as Fig.\ \ref{fig:PLYDD}
and \ref{fig:PLYDDEvary} show.

Given this exact solution for the degree distribution, its
generating function
\beq
 G(z) := \sum_{k=0}^{E} n(k) z^k ,
\eeq
may be obtained exactly in terms of the hypergeometric function
$F(a,b;c;z)$:
\bea
 G(z) &=& n(0) F(\widetilde{K},-E;1+\widetilde{K}-E-\widetilde{E}; z) .
 \label{GenFuncSol}
\eea
The $m$-th moments of the degree distribution are then
\beq
\frac{1}{G(1)} \! \left. \frac{d^mG(z)}{dz^m}\right|_{z=1}
 \! =
 \frac{\Gamma(\widetilde{K}+m)\Gamma(-E+m)\Gamma(1-\widetilde{E}-m)}
      {\Gamma(\widetilde{K})\Gamma(-E)\Gamma(1-\widetilde{E})}
\eeq
In particular the case $m=0$ fixes the normalisation, $A$, of $n(k)$
in \tref{nsimpsol}, while $m=1$ confirms the results are completely
consistent in the determination of $\kav$.

\begin{figure}
\includegraphics[width=7.8cm]{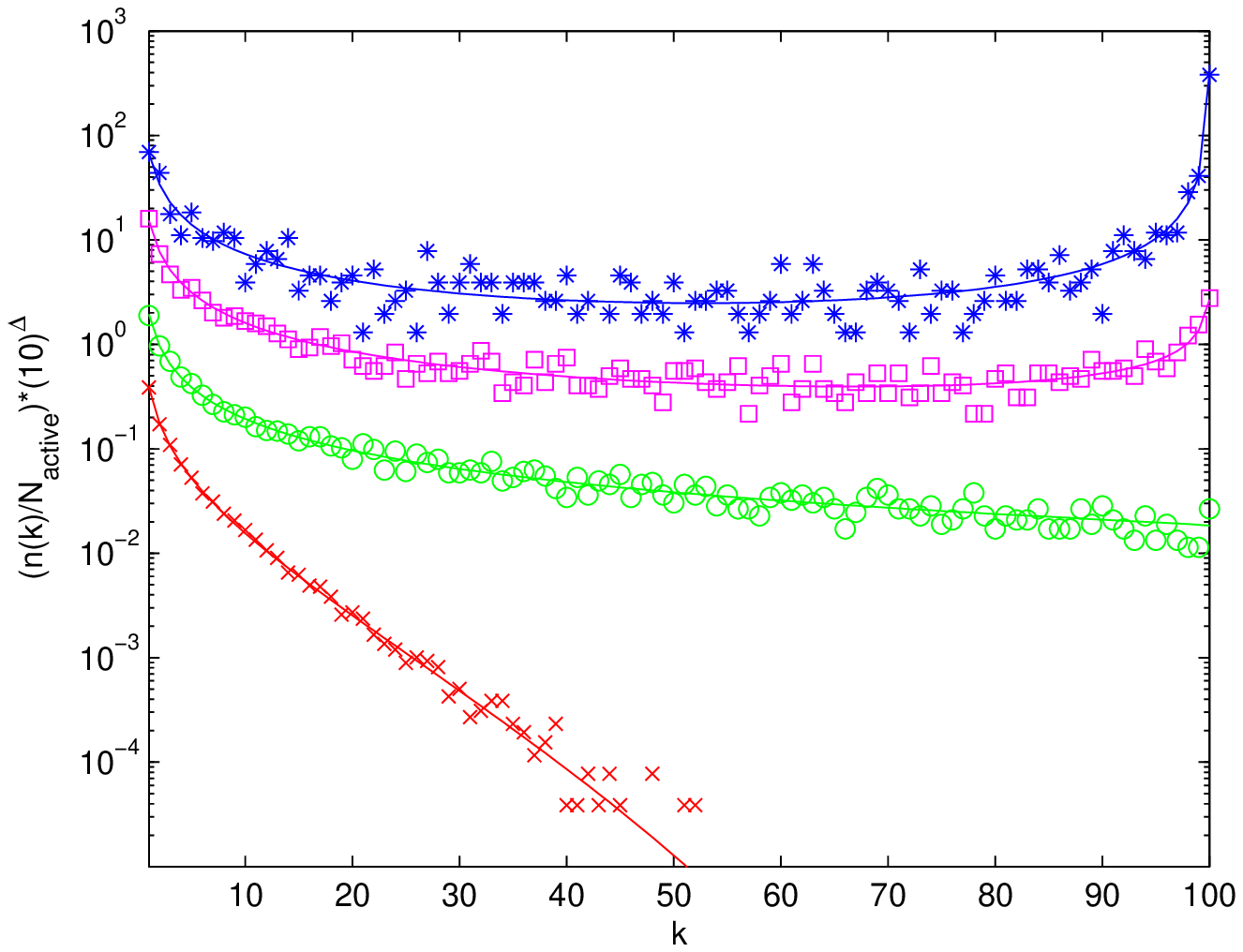}
\includegraphics[width=7.8cm]{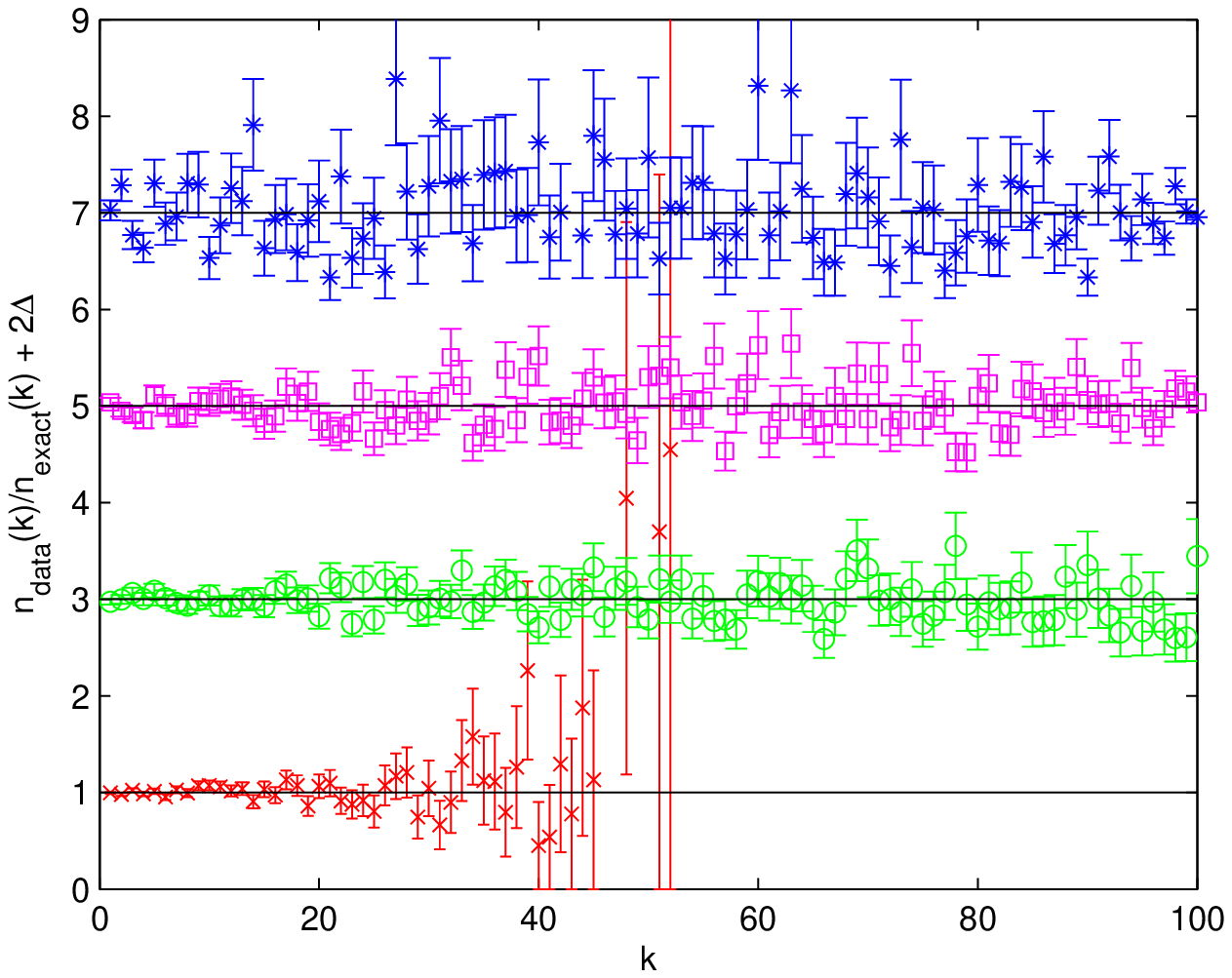}
\caption{Plots of the degree distribution, normalised by the sum of
values for degree greater than zero, and the fractional error of the
data w.r.t.\ the exact solution. For each parameter set plots
shifted by a constant controlled by $\Delta$ for clarity. For
$E=100$ but with new artifacts added with probability
$\bar{p}=1-p_p$ ($p_r=0$) where $\bar{p}=0.1$ (crosses, $\Delta=0$),
$0.01$ (circles, $\Delta=1$), $0.005$ (squares, $\Delta=2$) and
$0.001$ (stars, $\Delta=3$). The lines are the relevant equivalent
mean field solutions. Measured after $10^5$ rewiring events, and
averaged over $10^4$ runs.  Started with $n(k=1)=E$ but otherwise
$n(k)=0$. Errors on the degree distribution are not shown.}
\label{fig:BHSDD}
\end{figure}

There is another important attachment process that may be included
in this model. Suppose that with probability $\bar{p}=1-p_r-p_p$ a
new artifact vertex is added to the network. The new artifact
receives the edge removed from an existing artifact on the same time
step. The cultural transmission models
\cite{Neiman95,BSHB03,BHS04,HBH04} and the gene pool model studied
in \cite{KC64,CK70} include this process. In the long time limit the
number of artifacts becomes infinite, and the random attachment then
becomes completely equivalent to this process of new artifact
addition. This is the large $N$, zero $\kav$ limit of the discussion
above.  Care is needed as $n(0)$ diverges and an alternative
normalisation is needed. The degree distribution for $k \geq 1$
behaves in exactly the same way as before, a simple inverse degree
power law cutoff by an exponential for $E(1-p_p) \gtrsim 1$ but for
$E(1-p_p) \lesssim 1$ a single artifact is chosen by most
individuals. Intriguingly for this model when $E(1-p_p) = p_p$ the
degree distribution an exact inverse power law for the \emph{whole}
range of non-zero degrees.  The exact solution to the mean field
equations again provides an excellent fit to the data as Fig.\
\ref{fig:BHSDD} shows.

These results have several implications. First I have noted that
many apparently different models are all equivalent to this simple
bipartite network model.  Then in terms of mathematical detail,
previous mean field equations did not include the $(1-\Pi)$ terms of
\tref{neqngen}.  Thus exact solutions given here are novel. The
various forms for the asymptotic behaviour found in the literature
can now be seen to be various small $p_r$ and/or large $E$
approximations to the exact results, e.g.\ descriptions elsewhere of
the condensation regime $p_rE \ll 1$ are for large $E$. Further the
calculation of the generating function shows that all aspects of
this model appear to be analytically tractable so this rewiring
model may prove to be as useful the Erd\H{o}s-R\'eyni random graph.

As noted in the introduction, the model also has a wide range of
practical applications
\cite{Neiman95,BSHB03,BHS04,HBH04,KC64,CK70,Ritort95,BBJ99,BDJKNPZ02}.
While it may be too simple in practice, it does at worst give a
useful null model against which to test other hypotheses.

However copying the choice of others could also be a genuine
strategy, even if it emerges as the result of a more fundamental
process. Suppose that the individuals are connected to each other by
a second network.  When making the choice of artifact for
attachment, the individual which is rewiring its edge could consult
its acquaintances as represented by this network, and may well
choose to follow their recommendation, i.e.\ copy their artifact.
Such random walks on a network, even when of length one, lead
naturally to the emergence of preferential attachment in most cases,
\cite{RndWalk}. This explains results in a model of the Minority
Game \cite{ATBK04}. There individuals are connected by a random
graph and choose a strategy (the artifact) by copying the `best' of
their neighbours. If what is best is continually changing then for
the degree distribution this will be statistically equivalent to
copying the strategy of a random individual.  It is no surprise then
that the results for the popularity of strategies in \cite{ATBK04}
follows a simple inverse power law with a large degree cutoff.

Finally one may consider the scaling properties of the model. In
examples such as pottery styles or dog breeds, the categories
assigned by investigators are a coarse graining imposed on a
collection where each individual is really unique at some level.
However one would hope that the results are largely independent of
this categorisation. So suppose the artifacts are paired off at
random. The decision to copy or to innovate on a given event do not
change, so $p_r,p_p$ and $\bar{p}$ remain the same.  Because
preferential attachment is \emph{linear} in degree, the probability
of preferentially attaching to a given pair of artifact vertices is
just proportional to the sum of their degrees which in turn is just
the degree of the artifact pair vertex.  Thus we retain preferential
attachment.  The probability of choosing one of a pair of artifacts
at random is double choosing just one at random but this reflects
that the number of artifact pairs $N_2=N/2$ is just half the
original number of artifact vertices. Overall, the form of the
equations for the degree distribution of these pairs, $n_2(k)$, is
exactly as before and the only parameters which change are $N
\rightarrow N/2$ and $\kav \rightarrow \kav /2$.  Thus the generic
form of the distribution of artifact choice is independent of how
artifacts are classified though the detailed prescription changes in
a simple manner.

I thank D.Brody, H.Morgan, A.D.K.Plato and W.Swanell for useful
conversations.

% ******************************************************************
% The Bibliography

\end{document}